\documentclass[a4paper]{jpconf}
\usepackage{iopams,amsfonts,amsmath,amssymb}
\usepackage{calrsfs}
\usepackage[T1]{fontenc}
\usepackage{textcomp}
\usepackage{times}
\usepackage{bm}
\usepackage{iopams}
\usepackage[dvips]{color,graphicx}
\usepackage[dvipsnames]{pstricks}
\usepackage{url}
\usepackage{upgreek}
\usepackage[square,comma,sort&compress,numbers]{natbib}
\usepackage[ps2pdf,colorlinks=true,linkcolor=blue,anchorcolor=blue,runcolor=blue,urlcolor=blue,citecolor=blue,unicode=true,pdfdisplaydoctitle=true]{hyperref}
\newcommand{\REM}[1]{}
\eqnobysec
\begin{document}
\noindent{\blue\sffamily To appear in Journal of Physics: Conference Series}

\title{Interaction anisotropy and random impurities effects on the critical
behaviour of ferromagnets}
\author{H Chamati $^1$ and S Romano $^2$}
\address{$^1$ Institute of Solid State Physics,
Bulgarian Academy of Sciences,
72 Tzarigradsko Chauss\'ee, 1784 Sofia, Bulgaria}
\ead{chamati@issp.bas.bg}

\vspace*{.2cm}

\address{$^2$ Dipartimento di Fisica ``A. Volta``,
Universit\`a di Pavia, Via A. Bassi 6, I-27100 Pavia, Italy}
\ead{Silvano.Romano@pv.infn.it}

\begin{abstract}
The theory of phase transitions is based on the consideration of
"idealized" models, such as the Ising model: a system of magnetic moments
living on a cubic lattice and having only two accessible states. For
simplicity the interaction is supposed to be restricted to
nearest--neighbour sites only. For these models, statistical physics
gives a detailed description of the behaviour of various thermodynamic
quantities in the vicinity of the transition temperature. These findings
are confirmed by the most precise experiments. On the other hand,
there exist other cases, where one must account for additional
features, such as anisotropy, defects, dilution or any effect that
may affect the nature and/or the range of the interaction. These
features may have impact on
the order of the phase transition in the ideal model or smear it
out. Here we address two classes of models where the nature of the
transition is altered by the presence of anisotropy or dilution.
\end{abstract}

\section{Introduction}
Materials in Nature can be found in qualitatively different phases
having distinct properties. The change from one phase to another is
the consequence of a variation of an intensive thermodynamic quantity,
e.g., the temperature $T$, the pressure $P$,
external electric or magnetic
fields $E$ or $H$. Phase transitions are accompanied by abrupt changes in a
number of macroscopic thermodynamic quantities. Some familiar examples of phase
transitions include the gas--liquid transition (condensation), the
liquid--solid transition (freezing), the normal--superconducting
transition in conductors, the paramagnet--ferromagnet transition in
magnetic materials, and the superfluid transition in liquid helium.
Further examples are transitions involving amorphous or glassy
structures, spin glasses, liquid crystals, charge--density waves, and
spin--density waves.

In many cases, the two phases above and below the 
transition point, say $T_c$ for a
temperature driven phase transition, may be
discerned from each other in terms of some ordering
that occurs in the phase below $T_c$.
For example in
the liquid--solid transition the molecules of the liquid get
``ordered'' in space when they form the solid phase. 
In a paramagnet, the magnetic
moments of each atom can point in any random direction (in the
absence of an external magnetic field), but in the ferromagnetic phase
the moments are lined up in a particular direction of ordering.
Thus in the high temperature phase (above $T_c$), the
degree of ordering is smaller than in the low temperature phase (below
$T_c$). To quantify the amount of ordering in a system one uses the so called
\textit{order parameter}, which is usually a vanishing quantity in the
high temperature (disordered) phase.

The phase diagram shows regions within which homogeneous 
equilibrium states exist as a function of temperature and other thermodynamic
variables like $P$, $E$,  $H$. For some physical systems the chemical potential $\mu$
or composition variables are also involved.
The different regions of the phase diagram are delimited by phase
boundaries that mark conditions under which multiple phases can
coexist at equilibrium. Phase transitions take place along phase
boundaries marked by lines of equilibrium.

The theoretical framework
that aims at describing phase transitions
and related phenomena as a result from cooperative effects over macroscopic
scales is a part of the realm of equilibrium statistical physics
\cite{yeomans1992,pathria1996,mazenko2003}.
Statistical physics is based on
probabilistic models of the interactions of microscopic entities
forming large assemblages (macroscopic bodies).
The probability a macroscopic system of volume $V$, having $N$
particles, in a state $\mathfrak{S}$ with energy
$E$ at a temperature $T$, is given by
\begin{equation}\label{gibbs}
\mathcal{P}(\mathfrak{S})=\frac{e^{-E(\mathfrak{S})/(k_B T)}}{\mathcal{Z}(T)},
\end{equation}
where $k_B$ is the Boltzmann constant, and
$\mathcal{Z}(T)$ is the partition function
\begin{equation}
\mathcal{Z}(T)=\sum_{\mathfrak{S}}e^{-E(\mathfrak{S})/(k_BT)},
\end{equation}
relating microscopic degrees of freedom to macroscopic thermodynamic
quantities. The function $\mathcal{Z}(T)$ depends upon any parameters that
might affect the value of $E(\mathfrak{S})$. The expectation value of
any statistical operator $\mathfrak{O}$ is defined via
\begin{equation}
\langle\mathfrak{O}(\mathfrak{S})\rangle=
\sum_{\mathfrak{S}}\mathfrak{O}(\mathfrak{S})\mathcal{P}(\mathfrak{S}).
\end{equation}
For example the internal energy $U$ is obtained by averaging
$E(\mathfrak{S})$ over
all the accessible states of the system. This is given by
$$
U=\langle E(\mathfrak{S})\rangle.
$$

A fundamental thermodynamic quantity related to the partition function
is the \textit{free energy}
\begin{equation}
f(T)=-k_B T \ln \mathcal{Z}(T),
\end{equation}
which contains all the
information on the thermodynamics of the considered system.

According to the Ehrenfest classification scheme there are different kinds of
phase transitions depending on the nature of the singularities of the
thermodynamic quantities at the transition. Such transitions are
categorized as first--, second--, or higher order transitions if the
lowest derivative of the free energy that exhibit
nonanalytic behaviour with a finite jump
 is the first, second or higher one. Within this
classification the Berezinski\v\i--Kosterlitz--Thouless (BKT)
\cite{berezinskii1971,kosterlitz1973}
may be considered as being of infinite order.


To describe universal features of phase transitions one uses
relatively ``simple'' microscopic models such as the magnetic model
involving interaction between magnetic degrees of freedom.
In a \textit{classical model} for a magnet, the spins (magnetic
moments) may be represented by
$n$--component unit vectors $\bm{s}_i$ located at sites $i$, with
coordinates $\bm{x}_i$, 
belonging to a finite subset of a generic $d$--dimensional lattice $\Lambda_d$,
i.e.  
$\mathfrak{L}\subset\Lambda_d$, with
$|\mathfrak{L}|=N$. We are considering
here saturated lattice models, where each lattice site hosts
one spin.
The simplest and probably most extensively studied cases considered in the 
literature assume a hypercubic
lattice $\Lambda _d = \mathbb Z^d$ and 
isotropically interacting spins, thus the (nearest--neighbour) Hamiltonian
\begin{equation}
\label{heisenberg}
\mathcal{H}_\mathfrak{L}=-\frac J2\sum_{\langle i,j\rangle}
\bm{s}_i\cdot\bm{s}_j-\bm{H}\sum_{i}\bm{s}_i,
\end{equation}
where the coupling $J$ is restricted to
nearest--neighbouring sites $i$ and $j$,
\REM{\footnote{This is a very good approximation if one takes into account
screening effects and multipolar interactions, caused by electric
charges and currents of opposite signs, that tend to reduce the
range of the interaction making it effectively short ranged.}}
with each distinct pair being counted once. 
\REM{\footnote{\textcolor{blue}{The unit vectors $\bm{s}_i$ shall be
parameterized by the usual polar angles $\varphi_i$ ($n=2$)
or spherical ones  $\theta_i$ and  $\phi_i$ ($n=3$),
i.e. 
$\bm{s}_i=(\cos\varphi_i,\sin\varphi_i)$, or 
$\bm{s}_i=(\sin\theta_i\cos\phi_i,\sin\theta_i\sin\phi_i,\cos\theta_i)$
.}}}
$\bm{H}$ stands for an uniform magnetic field.

In the absence of the magnetic field, i.e. $\bm{H}=\bm{0}$, a
ferromagnetic coupling, $J>0$,
favours a parallel orientation of the
spins in the ground state, whereas thermal  fluctuations tend to
create an orientational disorder. On the other hand an
interaction, $J<0$, would be the precursor of an
antiferromagnetic order at low temperatures.
Actually, in the specific case considered
here, i.e. nearest--neighbour coupling and bipartite lattice,
and with $\bm{H}=\bm{0}$, 
the sign of the coupling constant is immaterial, i.e.
models defined by $+J$ and $-J$ yield the same partition
function, whereas the two correlation functions
are connected by suitable sign factors.

For $n=1,2,3$, the model corresponds to the Ising, planar rotator (PR)
and Heisenberg (He), respectively. The Ising model is known to have a
$Z_2$ discrete symmetry, while systems
with $n\geq2$, like PR and He, are said to possess a
continuous $O(n)$ symmetry.
It has become customary to refer to the number of component $n$ of
the order parameter as symmetry index.
The interaction with an external field breaks
this symmetry and establishes a preferred direction for spin
alignment. By reducing the external field to zero in the thermodynamic
limit, the system may exhibit spontaneous magnetization pointing in
the initial direction of the field.
It has been shown \cite{stanley1968} that the limit
$n\to\infty$ leads to the spherical model \cite{berlin1952}
obtained by requiring the spins to be continuous variables subject to
a global relaxed constraint ($\sum_{i=1}^N\bm{s}_i^2=N$) rather
than forcing them to take unit lengths i.e. $|\bm{s}_i|^2=1$.
The formal limit $n\to0$ is relevant to the study of self--avoiding walk
problem, which can be applied to polymers.

By now a number of rigorous results, assuming translational
invariance, have been
worked out, entailing existence or absence of a phase transition in
the thermodynamic limit,
depending on lattice dimensionality $d$ and number of spins components
$n$ \cite{georgii1988,sinai1982}.

Model (\ref{heisenberg}) with different values of $n$ and $d$
is extensively studied in the literature via different methods
and its behaviour as a function of the temperature is very well
known. For a review with a rich list of references see
\cite{pelissetto2002}. At $H=0$,
it exhibits a second order phase transition,\footnote{Here and below the
temperature will be measured in units of $J/k_B$.}
for any $d>1$ and with discrete spin
variables, i.e. $n=1$. Such a transition is characterized by
a significant growth of the nearest--neighbour correlations
for orientational fluctuations, and also
the onset for long--range orientational correlations.
On the other hand, according to the Mermin--Wegner theorem \cite{mermin1966}
in $O(n)$ symmetric
models ($n\geq2$) there can be no spontaneous symmetry breaking at finite
temperatures for
$d\leq2$ meaning that the system remains orientationally
disordered at any finite temperature.
For $d>2$ and $n\geq2$ a second order phase
transition takes place in the system.
In the two--dimensional case $d=2$ and $n=2$ the system
exhibits a BKT transition from a high temperature disordered phase to
a low temperature phase with slow decay of the correlation function and an 
infinite magnetic susceptibility \cite{gulacsi1998}.

In the vicinity of a continuous phase transition, such as second order
or
BKT transition, there is only one dominating length scale related to
the growth of fluctuations: The \textit{correlation length}.
Because of the diverging nature of the correlation length as the critical
point is approached the microscopic details of the system
becomes irrelevant. Thus the description of
the singular behaviour of many thermodynamic observables requires a
small number of universal variables: critical exponents, amplitudes
and functions. This allows the arrangement of a great
variety of different microscopic systems in universality classes of
equivalent critical behaviour. A universality class depends upon the
number of components of the order parameter and the dimensionality of
the system. For more details the reader is invited to consult
references \cite{fisher1998,stanley1999}.

The model (\ref{heisenberg}) can describe the properties
of a wide variety of physical
situations in the vicinity of their transition point, 
however, it may happen that the very experimental situation at hand 
requests the interaction model to be complicated  
(made physically richer and theoretically
more challenging) in various ways.
On the one hand, there exist different possible lattice types
$\Lambda_d$, in addition to $\mathbb Z^d$. 
On the other hand, as for the orientational dependence,
more elaborate potential models involve 
(in some combination or other) isotropic or anisotropic linear 
couplings between spin components, sometimes higher powers of scalar 
products among the interacting spins,
multipolar (usually dipolar) interactions, Dzyaloshinski--Moriya
terms, single--site anisotropy fields.
Notice also that more distant neighbours are sometimes involved,
or even, in principle, all neighbours may be coupled 
by long--range interactions.
In some specific favourable cases one has even been able to match the
model and its potential parameters to a specific experimental system
\cite{dejongh2001}.
The ferromagnetic  ordering transition 
observed experimentally in the absence of an external
field is more frequently second order, but first--order
transitions are also known. They  might, for example,  result from doping 
by nonmagnetic impurities, 
anisotropy of interactions in  spin space, or coupling to the
lattice \cite{binder1987}.
Other interaction models  for different physical systems can be found in
\cite{cardy1996}.

This review is devoted to the description and analysis of effects
related to the nature of the interaction.
This aims
at gaining insights in the thermodynamics of some specific systems. The models
considered here are some of the most popular models in the theory of
phase transitions. They involve
different kind of interactions that hopefully might be adapted to the 
characteristics of a given material. They
describe ferromagnetism with anisotropic coupling,
systems with random dilution and may 
to some extent be used to investigate fluids.
Special attention is paid to the construction of the phase diagrams of 
these models
that are determined from the investigation of
different thermodynamic quantities such as: the free energy, the
susceptibility, specific heat etc.

The review is organised as follows: In Section \ref{xymodels} we
discuss the transitional behaviour of \textit{generalised XY} models
introduced in reference \cite{romano2002}.
These are generalization of the XY model with a nontrivial
coupling along the $z$ components of the spins. We construct the phase
diagrams of the models in two and three dimensions and determine the
effect of the nontrivial coupling on the nature and location of
transition temperature. In Section \ref{dilution} we review the effect
of dilution of ferromagnets by the introduction of random impurities
and present the phase diagram of the diluted Heisenberg in three dimensions
and the diluted plane rotator in two dimensions. We conclude with
Section \ref{concl}, where we discuss the results and comment on
other models sharing similar transitional behaviour.

\section{Generalised XY models}\label{xymodels}
Let us first consider
interactions being anisotropic in spin space with nonvanishing
and equal ferromagnetic couplings involving $m<n$ components 
of the partner spins only, while still
keeping the interaction restricted to nearest neighbours. In this case
the model reads
\begin{equation}
\label{anisoeq}
\mathcal{H}_\mathrm{anis.}= -\frac J2\sum_{\langle
i,j\rangle}\sum_{\alpha,\beta\leq m}
\delta_{\alpha,\beta} s_i^\alpha s_j^\beta.
\end{equation}
Among these models we may mention the \textit{continuous} Ising
($m=1$ and $n$ arbitrary) and
the various versions of XY models ($m=2$ and $n$ arbitrary),
where the interaction is restricted to only one component and
two components of the magnetic moments, respectively

In general the transitional behaviour of such models is analogous to
that of their $O(m)$ isotropic counterparts:
On the one hand, the universal critical behaviour of 
these models in the vicinity of their
corresponding transition temperature is equivalent to their isotropic
$O(m)$ counterparts i.e. the models share the same
universal features (critical exponents and amplitudes).
on the other hand, the transition temperature
is a typical non--universal quantity, and is recognizably affected
by the anisotropy.

A more general class of anisotropic spin models may be constructed by
introducing some kind of extreme anisotropy coupling only
a part of the spin components in some nontrivial manner in addition to
the anisotropic interaction in the spin space.
A model that will
be discussed in this study is the so called ``generalised'' XY model introduced
in reference \cite{romano2002}. This model involves 
$3-$component unit vectors, and is defined by
\begin{equation}\label{genXYm}
\mathcal{H}_{XY}^p=-\frac J2\sum_{\langle i,j \rangle}
(\sin\theta_{i}\sin\theta_{j})^{p}\cos(\phi_{i}-\phi_{j}) ,
\end{equation}
where $p \in \mathbb{N}$
is a parameter controlling the strength of anisotropy along the
$z$--spin direction,
and the spins are expressed in terms of the usual spherical
coordinates $\theta_i$ and $\phi_i$ i.e.
$\bm{s}_i=(\sin\theta_i\cos\phi_i,\sin\theta_i\sin\phi_i,\cos\theta_i)$.
Notice that by setting
$p=1$ ($p=0$) we recover the familiar XY model (planar rotator).
In this case of the planar rotator model, the
$\theta$--dependence only survives in the free--spin measure.
The Hamiltonian (\ref{genXYm})
can also be written in terms of spin components,
in the more complicated form
\begin{equation}
\label{RZ}
\mathcal{H}_{XY}^p=-\frac J2\sum_{\langle i,j \rangle}
\left[1-(s_{i}^{z})^{2}-(s_{j}^{z})^{2}+(s_{i}^{z}s_{j}^{z})^{2}\right]^{(p-1)/2}
\left(s_i^x s_j^x + s_i^y s_j^y \right).
\end{equation}
As for the role of $p$ in Eq. (\ref{genXYm}),
notice that it could be taken 
to be a real positive number, say ranging between $0$ and $1$
(and hence continuously interpolating between planar rotator and XY models).
On the other hand, larger values of $p$ reinforce
the out--of--plane fluctuations.
This makes it possible to widely vary the anchoring of spins 
with respect to the horizontal plane
which might have direct experimental relevance.
As for the present model, this change of anchoring
is ultimately reflected by the significant changes in transition 
behaviour.

The transitional behaviour of the model in two and three dimensions has been
investigated by different approaches. First, it was
proven rigorously \cite{romano2002}, on the basis of the known
behaviour of PR, 
that when $d=2$ and for all values of $p$,
the named potential models produce orientational disorder
at all finite temperatures, and support a BKT--like transition.
On the other hand, when $d=3$, 
these models support ordering transitions taking place at
finite temperatures. In both cases, the transition temperatures
are bounded from above by the corresponding values for the PR
counterpart. It was later proven \cite{vanenter2006},
again rigorously,
that the transition turns first--order for sufficiently large $p$.
Notice that the threshold values had to be estimated by
other means.

\subsection{Two dimensions}
Using Monte Carlo simulations, the thermodynamics of model
(\ref{genXYm}) has been investigated in details in reference
\cite{mol2006} for $d=2$ and values of $p$ ranging from 2 to 5.
Analysis of
Monte Carlo data, showed that the model
produces a BKT--(like) transition, possibly changing
to a first--order transition for larger $p$,
due to the large number of vortices and strong
out--of--plane fluctuations.
It has been found that the transition
temperature is indeed decreasing with increasing $p$.

To gain insight into the transitional behaviour of the model for
large values of $p$ we performed further Monte Carlo simulations at
$p=6$.
Our analysis shows that the transition is most likely to have a weak
first order nature. Unfortunately more simulations are required and different
approaches are needed to be more conclusive.


\begin{figure}[ht!]
\centering
\resizebox{0.5\columnwidth}{!}{\includegraphics{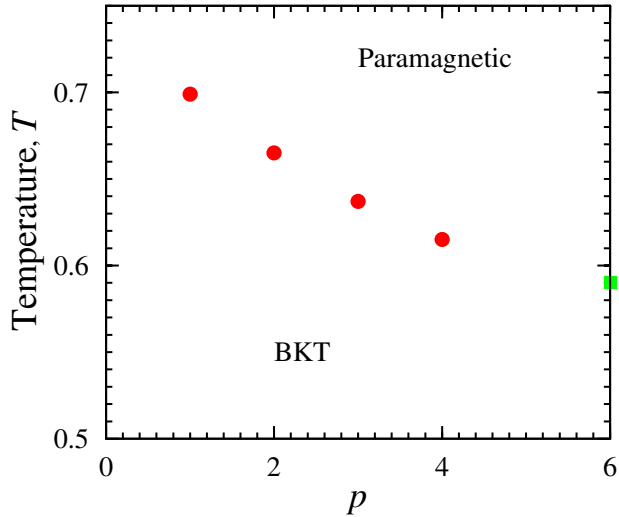}}
\caption{($p,T$) phase diagram of the two--dimensional generalised XY
model. Full circles indicate a Berezinski\v\i--Kosterlitz--Thouless
like phase transition, while the square is believed to corresponds to
a weak first order phase transition.}
\label{diagram2dxy}
\end{figure}

In figure \ref{diagram2dxy} we show the phase diagram of the two
dimensional generalised XY model in the $(p,T)$ plane. It is seen that the model
exhibits a phase transition from a paramagnetic phase to a BKT--like
phase as the temperature decreases. The transition temperature is
evaluated using Monte Carlo simulations according to reference
\cite{mol2006}. The value of the transition temperature corresponding
to $p=1$ is taken from reference \cite{evertz1996}.

\subsection{Three dimensions}

For $d=3$, different analytical approximations,  such as
a Mean Field (MF) approach, as well as its Two--Site Cluster (TSC) 
refinement have been used in reference \cite{romano2002} to estimate 
transition temperatures for $p=2,3,4$, and then
for higher values of $p$ in subsequent papers (see below).
Notice also
that $(\sin \theta)^p$, and hence the absolute value of the
interaction potential, decreases with increasing $p$, and this aspect
is reflected by the $p$--dependence of the estimated transition temperature.
Transition temperatures have been estimated in reference \cite{mol2003}
by self--consistent harmonic approximation,
both for $d=2$ and $d=3$, and it was found that the
transition temperature is decreasing against $p$. A study
of the model in its continuum limit, carried out in reference
\cite{mol2003} also showed that out--of--plane fluctuations,
and consequently the magnon density, decrease
with increasing $p$. 

We have also addressed the three dimensional generalised XY models
for various values of $p$, by means
of Monte Carlo (MC) simulation.
We have investigated the transitional behaviour of
various thermodynamic functions, such as the susceptibility and
the specific heat, and made comparisons with MF and TSC
predictions.
MF  yielded a tricritical behaviour with
tricritical points having real (non--integer)
values of the parameter $p$. As for simulation results,
transitional behaviour characetristic of the XY model was found for
$p=2,3,4,8$,
the case $p=12$ suggested tricritical behaviour,
whereas evidence of first--order transitions  was obtained for $p=16,20$.

\begin{figure}[ht!]
\centering
\resizebox{0.5\columnwidth}{!}{\includegraphics{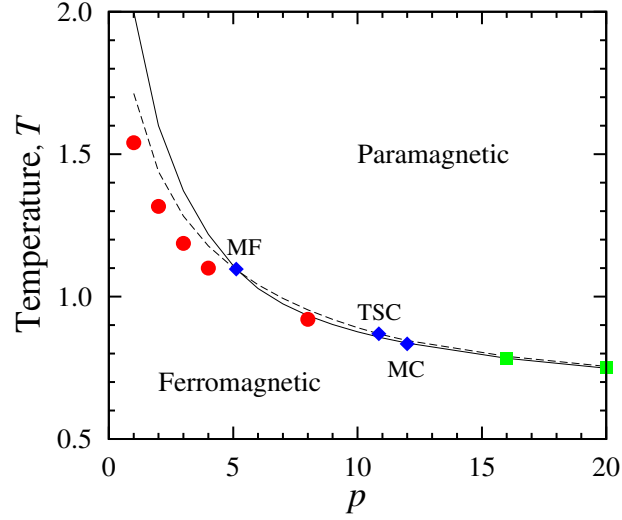}}
\caption{
Phase diagram of the three dimensional generalised XY model. The solid
and dashed lines stand for transition temperatures estimated via
mean--field and two--site--cluster treatments, respectively.
Full circles and squares indicate second and first order phase
transitions temperatures evaluated via Monte Carlo simulations,
respectively. Full diamonds shows the locations of tricritical points.}
\label{diagram3dxy}
\end{figure}

In Figure \ref{diagram3dxy} we present the $(p,T)$ phase diagram,
where we can read off the behaviour of the transition
temperature as a function of $p$ for $1\leq p\leq20$. MF and TSC data
can be found in references
\cite{romano2002,chamati2005c,chamati2006c,vanenter2006}, 
while Monte carlo data
are taken from references \cite{costa1996,chamati2005c,chamati2006c}.
Notice that, for $p \le 4$, TSC
gives better estimates of $T_c$ than MF, then
the two roles are exchanged for $p=8,12$, and finally the three methods
give very similar answers  when $p \geq 16$, i.e. where  the transition
has a pronounced first--order character.

\section{Annealed dilution in ferromagnets}\label{dilution}
Another class of systems that has been extensively studied in the
literature spans those  with random impurities (disorder). For a review on
diluted magnetism see reference \cite{stinchcombe1983}.
The interest in these systems stems
from the fact that in nature no system is really pure. Indeed,
the presence of cavities, grain boundaries, lattice defects,
chemical impurities (i.e. of  other chemical individuals) or some other kind of
disorder may affect the properties of the pure system,
and result in different effects.
To consider a simple but rather important example, 
in a system of magnetically coupled spins,
some lattice sites may be occupied by
nonmagnetic constituents (site--dilution). 
Both in experimental and theoretical terms, one can distinguish
between  annealed and quenched disorder.

In quenched systems, impurities are held frozen in randomly distributed 
fixed positions, without the possibility of overcoming  potential barriers 
for diffusing into the host material: in this case the relaxation time is 
very long and thermal equilibrium between impurities and the constituents of 
the host is never reached. 
In annealed materials, impurities are allowed to diffuse
randomly and to reach thermal equilibrium with the other constituents of 
the host material.
One can also think of a two--component solution being very dilute
with respect to one of the components. When the system is in the liquid phase,
molecules of the two types can exchange their positions, and diffuse
throughout the sample. Let then the system crystallize, at sufficiently
low temperature. In the resulting solid, particles of the
minority component are fixed in certain lattice sites only.

The simplest extension of equation (\ref{heisenberg}) taking these
situations into account reads
\begin{equation}
\label{Hp}
\mathcal{H}_\mathrm{dis.}=-\frac J2\sum_{\langle i,j\rangle}
\nu_i\nu_j\ \bm{s}_i\cdot\bm{s}_j.
\end{equation}
where the occupation numbers $\nu_i$ equal ``zero'' for a site $i$ 
hosting a nonmagnetic impurity and ``one'' for a  magnetic particle.
The density of magnetic particles in the system is defined via
$$
\rho=\frac1N\sum_i^N\langle\nu_i\rangle.
$$
Within this notation the pure system, i.e. without impurities, corresponds to $\rho=1$.

In the annealed case, Hamiltonian (\ref{Hp}) can be interpreted as describing
a two--component system consisting of interconverting ``real''
($\nu_i=1$) and ``ghost'', ``virtual'' or ideal--gas particles
($\nu_i=0$). Both kinds of particles have the same kinetic energy,
and the total number of particles equals the
number of available lattice sites.
In this case one
works in the
Grand--Canonical Ensemble
where the probability (\ref{gibbs}) for a configuration, now involving 
the occupation numbers $\{\nu_i\}$, as well as the spins $\{\bm{s}_i\}$, 
is defined by
\begin{equation}
\mathcal{P}\propto
\exp\left[-\beta\left(\mathcal{H}_\mathrm{dis.}-\mu\sum_i\nu_i\right)\right],
\end{equation}
where
$\mu$ denotes the excess chemical potential of ``real'' particles
over ``ideal'' ones.
Of course, the interaction may be anisotropic in spin space,
as in the cases outlined above, or involve more distant neighbours.
The system remains translationally invariant on average.
This model bears some similarity with the Blume--Emery--Griffiths
model \cite{blume1971} for $^3$He impurities in superfluid $^4$He. 
More precisely, notice also that, starting from an assigned model,
its lattice gas extensions can be written in general as
\begin{equation}\label{Hq}
\mathcal{H}_{LG}=-\frac J2\sum_{\langle i,j\rangle}
\nu_i\nu_j\ \bm{s}_i\cdot\bm{s}_j + 
\frac \lambda2 \sum_{\langle i,j\rangle} \nu_i\nu_j\ ,
\end{equation}
where the purely positional $\lambda$ term only becomes immaterial in
the pure limit
$\mu \rightarrow +\infty$. In equation (\ref{Hp}) we have chosen
the simplest case $\lambda=0$. Lattice gas models can be used to model
adsorption, and, in general, the variable
occupation numbers produce some fluidity of 
the system. One can also set the coupling $J$ term to zero,
so that the resulting model becomes isomorph with an
Ising model in external field. In general, the interplay
between $J$ and $\lambda$ can produce a richer phase diagram:
for example, when $\lambda>0$ and $\mu$ is sufficiently large,
the ground state may exhibit checkerboard positional
order but no orientational one.

It is very well known that a small
amount of annealed disorder does not affect the way the singularities
take place in pure systems i.e. the phase transition remains a second
order one. The $\rho$--dependent critical temperature is shifted towards
$T_c(\rho)<T_c(1)$.
If the chemical potential $\mu$ is held
fixed, the properties of the phase transition are the same as
those known for the pure system, corresponding to $\mu\to\infty$ and
$\rho=1$.
If, however, the concentration $\rho$
is kept fixed during the transition, care must taken in characterising
the transition. For details see reference \cite{fisher1968}.

A significant amount of impurities may alter the order of
the phase transition or make it disappear. Investigations of the XY
model, as a protype for He$^{3}$ -- He$^{4}$ mixtures, in three
dimensions, via high temperature series expansion
of the partition function, show that by increasing the 
density of randomness
the transition temperature decreases. At a certain value of $\rho$
the transition changes its nature and turns into a first order one
\cite{reeve1976}. The second order phase transition line
ends at a tricritical point, which marks the begining of a line of
fisrt--oder phase transition temperature that keeps decreasing as the
concentration of impurities increases.

In the three--dimensional case, the topology of the phase diagram of
model (\ref{Hp}) had been investigated by 
MF and TSC approximations for the Ising
\cite{sokolovskii2000}, as well as PR cases \cite{romano2000} in the
presence of a magnetic field, and for He at zero magnetic field
\cite{chamati2005a}. These investigations
were later extended \cite{chamati2005b} 
to the extremely anisotropic (Ising--like) two--dimensional model, and
in the absence of a magnetic field, as well.
The studied models were found to exhibit a
tricritical behaviour i.e. the ordering transition 
turned out to be  of first order
for $\mu$ below an appropriate 
threshold, and of  second order above it. When the transition is of 
first order, the orientationally ordered phase is also denser than the
disordered one.

To check the predictions of the
molecular--field--like treatments used to construct the phase
diagrams, extensive Monte Carlo simulations has been performed
\cite{romano2000,chamati2005a,chamati2005b} for particular 
values of the chemical potential.
A number of thermodynamic and structural properties had  been
investigated. It had  been found that there is a {\em second order} 
ferromagnetic phase transition 
manifested by a significant growth of magnetic and density
fluctuations. The transition temperatures were found to be
smaller than those of the corresponding values for the pure
systems and the critical behaviour
of the investigated models to be consistent with that of their pure
counterparts. Furthermore it had been found that MF yields a
qualitatively correct picture, and the quantitative agreement with
simulation could be improved by TSC, which has the advantage of
predicting two--site correlations. In general we found that simulations
results are consistent with the molecular--field like treatments.

\begin{figure}[ht!]
\centering
\resizebox{\columnwidth}{!}{\includegraphics{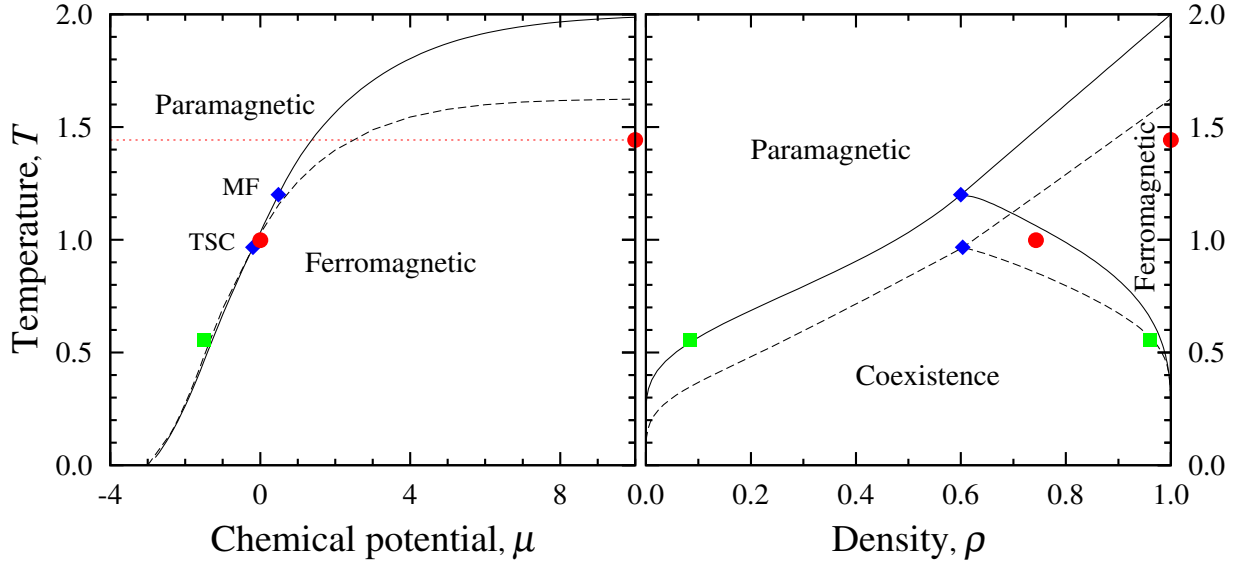}}
\caption{Phase diagrams of the three--dimensional Heisenberg
model in the chemical potential -- temperature and density -- temperature planes.
Solid and dashed lines correspond to mean--field and
two--site--cluster results, respectively. Circles and squares stand,
respectively, for estimates of the second and first order transition
temperatures using Monte Carlo simulations. Diamonds mark the
locations of tricritical points. The dotted line marks the
transition temperature of the model in the absence of impurities.
}
\label{diagramhe}
\end{figure}

In figure~\ref{diagramhe} we report the phase diagrams in the chemical
potential -- temperature and density -- temperature planes.
We show the behaviour of the transition temperature $T_c$
versus chemical potential $\mu$.
The plot also shows a fast, approximately linear,
increase of $T_c$ with $\mu$ up to $\mu \approx 0$, and a slower one
above this value. Moreover,  MF and TSC results
essentially coincide up to $\mu \approx 0$.
In the phase diagram in the ($\rho,T$) plane 
the existence of a first--order transition is reflected by a
biphasic region. We found that
the system exhibits a first--order phase transition from a dense
BKT phase to a paramagnetic one. In the temperature--density
phase diagram, both phases are expected to coexist over
some range of densities and temperatures.

Two--dimensional annealed lattice models were investigated
\cite{chamati2006e,chamati2007} as well, and the obtained results for $\mu=0$ or
a moderately negative $\mu$ were found to support a BKT phase
transition with a transition temperature lower than that of the pure parent due to
the presence of impurities. For negative and sufficiently large in magnitude
$\mu$ we found evidence of a first order phase transition in agreement with
renormalization group treatments \cite{cardy1979,berker1979}.

\begin{figure}[ht!]
\centering
\resizebox{\columnwidth}{!}{\includegraphics{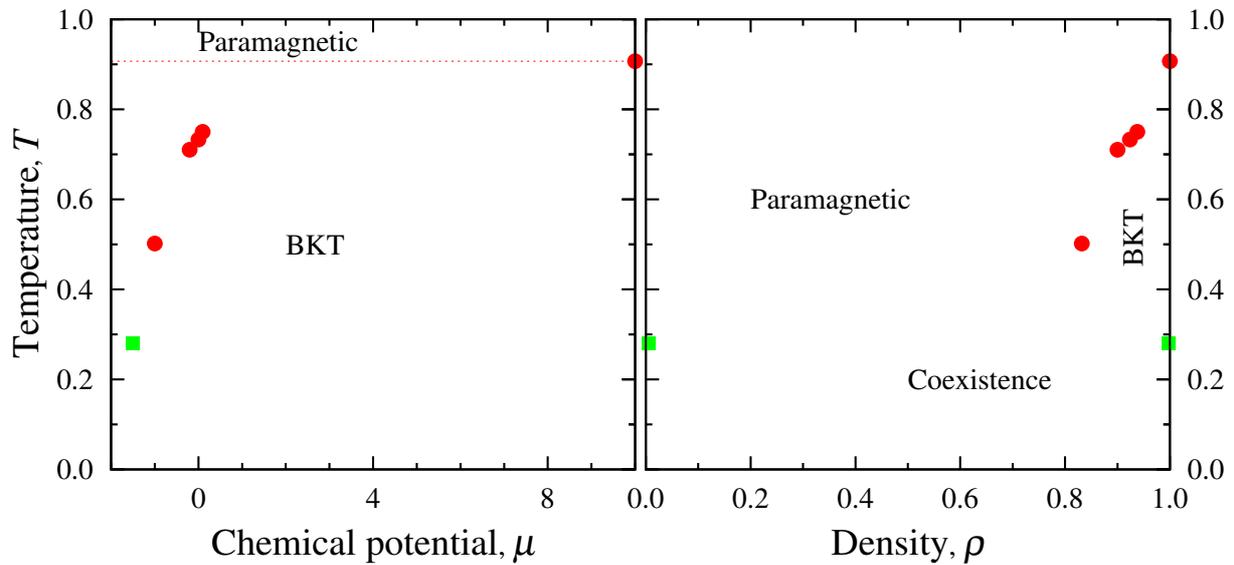}}
\caption{Phase diagrams of the two--dimensional planar rotator.
Circles and squares stand,
respectively, for estimates of the second and first order transition
temperatures using Monte Carlo simulations. Diamonds mark the
locations of tricritical points. The dotted line marks the
transition temperature of the model in the absence of impurities.
}
\label{diagrampr}
\end{figure}

In figure \ref{diagrampr} we show the phase diagram
in the $(\rho,T)$ plane. The existence of a first--order
transition is reflected here by a biphasic (binodal)
region. The phase diagram obtained here is
similar to the one resulting from the study of the
diluted planar rotator model \cite{romano2000}.

So far, we have discussed isotropic models. As we have mentioned, in
addition to dilution one can consider anisotropic interactions as well.
In reference \cite{chamati2005b} we have investigated two--dimensional continuous Ising spin
models with two and tree component spins. The phase diagrams of these
models were
found to be topologically similar to those found for the Heisenberg model
shown in figure \ref{diagramhe}. The main differences come from the
locations of the phase boundaries.

\section{Conclusion}\label{concl}
The Statistical Mechanics of lattice systems
is essentially based on the study of a number of relatively ``simple'' models
which are gradually made more complicated. We have shown here
some examples, involving generalized XY models and annealed magnetic
systems, where rigorous results have been produced to prove
existence and type of transition, supplemented by a variety of techniques
(MF or TSC approximation and Monte Carlo simulation) for elucidating the resulting physical
behaviour and estimating numerical values.

We obtained the phase diagram in two and three dimensions of a generalized version of the
lattice XY model, where the out--of--plane fluctuations of the spins are
controlled by a parameter $p$. Our investigation
shows that the nature of the transition is highly affected by the
strength of the out--of--plane fluctuations.
For $p\geq12$ at three dimensions and
$p\gtrsim6$ at two dimensions, it is first order,
whereas for small values its transitional behaviour
coincides with that of the original XY model.
The transition temperature is found to decrease
as $p$ increases at any dimension.

We constructed the phase diagrams of the annealed lattice plane
rotator in two dimensions and the Heisenberg model in three
dimensions. It is found
that the transition changes from a second order at three
dimensions and Berezinski\v\i--Kosterlitz--Thouless at two dimensions into a first
order transition as the concentration of impurities is increased.
In turn the transition temperature is found to decrease with
increasing impurity density.

\ack
This work was supported by the Ministry
of Education and Science of Bulgaria under Grant No $\Phi$--1517.


\begin{thebibliography}{10}
\expandafter\ifx\csname url\endcsname\relax
  \def\url#1{{\tt #1}}\fi
\expandafter\ifx\csname urlprefix\endcsname\relax\def\urlprefix{URL }\fi
\providecommand{\eprint}[2][]{\url{#2}}

\bibitem{yeomans1992}
Yeomans J~M 1992 {\em {Statistical Mechanics of Phase Transitions}\/} (New
  York: Oxford University \hspace*{.5cm} Press)

\bibitem{pathria1996}
Pathria R~K 1996 {\em {Statistical mechanics}\/} 2nd ed (Oxford, England:
  Butterworth--Heinemmann)

\bibitem{mazenko2003}
Mazenko G~F 2003 {\em {Fluctuations, Order and Defects}\/} (Hoboken: Wiley)

\bibitem{berezinskii1971}
{Berezinski\v\i} V~L 1971 {\em Sov. Phys. JETP\/} {\bf 32} 493--500 [Zh. Eksp.
  Teor. Fiz. 59, 907--920]

\bibitem{kosterlitz1973}
{Kosterlitz} J~M and {Thouless} D~J 1973 {\em J. Phys. C: Solid State Phys.\/}
  {\bf 6} 1181--1203

\bibitem{stanley1968}
Stanley H~E 1968 {\em Phys. Rev.\/} {\bf 176} 718--722

\bibitem{berlin1952}
Berlin T~H and Kac M 1952 {\em Phys. Rev.\/} {\bf 86} 821--835

\bibitem{georgii1988}
Georgii H~O 1988 {\em {Gibbs Measures and Phase Transitions}\/} ({\em de
  Gruyter Studies in Mathematics\/} \hspace*{.5cm} vol~9) (Berlin: Walter de Gruyter)

\bibitem{sinai1982}
Sinai Y~G 1982 {\em {Theory of Phase Transitions: Rigorous Results}\/} (Oxford:
  Pergamon)

\bibitem{pelissetto2002}
Pelissetto A and Vicari E 2002 {\em Phys. Rep.\/} {\bf 368} 549--727

\bibitem{mermin1966}
Mermin N~D and Wagner H 1966 {\em Phys. Rev. Lett.\/} {\bf 17} 1133--1136

\bibitem{gulacsi1998}
Gul\'acsi Z and Gul\'acsi M 1998 {\em Adv. Phys.\/} {\bf 47} 1--89

\bibitem{fisher1998}
Fisher M~E 1998 {\em Rev. Mod. Phys.\/} {\bf 70} 653--681

\bibitem{stanley1999}
Stanley H~E 1999 {\em Rev. Mod. Phys.\/} {\bf 71} S358--S366

\bibitem{dejongh2001}
{de Jongh} L~J and {Miedema} A~R 2001 {\em Adv. Phys.\/} {\bf 50} 947--1170

\bibitem{binder1987}
Binder K 1987 {\em Rep. Prog. Phys.\/} {\bf 50} 783--859

\bibitem{cardy1996}
Cardy J 1996 {\em {Scaling and Renormalization in Statistical Physics}\/} ({\em
  Cambridge Lecture Notes in \hspace*{.5cm} Physics\/} vol~5) (Cambridge, England: Cambridge
  University Press)

\bibitem{romano2002}
Romano S and Zagrebnov V 2002 {\em Phys. Lett. A\/} {\bf 301} 402--407

\bibitem{vanenter2006}
van Enter A~C~D, Romano S and Zagrebnov V~A 2006 {\em J. Phys. A: Math. Gen.\/}
  {\bf 39} L439--L445

\bibitem{mol2006}
M\'ol L, Pereira A~R, Chamati H and Romano S 2006 {\em Eur. Phys. J. B\/} {\bf
  50} 541--548

\bibitem{evertz1996}
Evertz H~G and Landau D~P 1996 {\em Phys. Rev. B\/} {\bf 54} 12302--12317

\bibitem{mol2003}
M\'ol L~A~S, Pereira A~R and Moura-Melo W~A 2003 {\em Phys. Lett. A\/} {\bf
  319} 114--121

\bibitem{chamati2005c}
Chamati H, Romano S, {M\'ol} L and Pereira A~R 2005 {\em {Eur. Phys. Lett.}\/}
  {\bf 72} 62--68

\bibitem{chamati2006c}
Chamati H and Romano S 2006 {\em Eur. Phys. J B\/} {\bf 54} 249--254

\bibitem{costa1996}
{Costa} B~V, {Pereira} A~R and {Pires} A~S~T 1996 {\em Phys. Rev. B\/} {\bf 54}
  3019--3021

\bibitem{stinchcombe1983}
Stinchcombe R~B 1983 {Dilute Magnetism} {\em {Phase Transitions and Critical
  Phenomena}\/} vol~7 ed \hspace*{.5cm} Domb C and Lebowitz J~L (London: Academic Press)
  chap~3, pp 151--280

\bibitem{blume1971}
Blume M, Emery V~J and Griffiths R~B 1971 {\em Phys. Rev. A\/} {\bf 4}
  1071--1077

\bibitem{fisher1968}
Fisher M~E 1968 {\em Phys. Rev.\/} {\bf 176} 257--272

\bibitem{reeve1976}
Reeve J~S 1976 {\em J. Phys. C: Solid State Phys.\/} {\bf 9} 2575--2587

\bibitem{sokolovskii2000}
Sokolovskii R~O 2000 {\em Phys. Rev. B\/} {\bf 61} 36--39

\bibitem{romano2000}
Romano S and Sokolovskii R~O 2000 {\em Phys. Rev. B\/} {\bf 61} 11379--11390

\bibitem{chamati2005a}
Chamati H and Romano S 2005 {\em Phys. Rev. B\/} {\bf 72} 064424

\bibitem{chamati2005b}
Chamati H and Romano S 2005 {\em Phys. Rev. B\/} {\bf 72} 064444

\bibitem{chamati2006e}
Chamati H and Romano S 2006 {\em Phys. Rev. B\/} {\bf 73} 184424

\bibitem{chamati2007}
Chamati H and Romano S 2007 {\em Phys. Rev. B\/} {\bf 75} 184413

\bibitem{cardy1979}
Cardy J~L and Scalapino D~J 1979 {\em Phys. Rev. B\/} {\bf 19} 1428--1436

\bibitem{berker1979}
Berker A~N and Nelson D~R 1979 {\em Phys. Rev. B\/} {\bf 19} 2488--2503

\end{thebibliography}

\providecommand{\newblock}{}

\centerline{\resizebox{0.0003\columnwidth}{!}{\includegraphics{void.eps}}}
\end{document}